\documentstyle[epsfig]{aipproc}

\begin{document}
\title{{\LARGE Radio limits on an isotropic flux \\
of $\ge 100$ EeV cosmic neutrinos}}

\author{P. W. Gorham$^*$, K. M. Liewer$^*$, C. J. Naudet$^*$, D. P. Saltzberg$^{\dagger}$, and D. R. Williams$^{\dagger}$}
\address{$^*$Jet Propulsion Laboratory, California Institute of Technology\\
4800 Oak Grove Drive, Pasadena, CA, 91109\\
$^{\dagger}$Dept. of Physics \& Astronomy\\
UCLA, Los Angeles, CA}

\maketitle

\begin{abstract}
We report on results from about 30 hours of livetime with the
Goldstone Lunar Ultra-high energy neutrino Experiment (GLUE).
The experiment searches for $\leq 10$ ns microwave pulses from
the lunar regolith, appearing in coincidence at two large
radio telescopes separated by about 22 km and linked by
optical fiber. The pulses can arise from subsurface 
electromagnetic cascades induced by interactions of up-coming
$\sim 100$ EeV neutrinos in the lunar regolith.
A new triggering method implemented after the first 12 hours
of livetime has significantly reduced the 
terrestrial interference background, and we now operate at the
thermal noise level. No strong candidates are yet seen. We
report on limits implied by this non-detection, based
on new Monte Carlo estimates of the efficiency. We
also report on preliminary analysis of smaller pulses,
where some indications of non-statistical excess may be
present.

\end{abstract}

\section{Introduction}
Recent accelerator results~\cite{Gor00,Sal01} 
have confirmed the 1962 prediction of Askaryan\cite{Ask62,Ask65}
that electromagnetic cascades in dense media should produce strong coherent
pulses of microwave Cherenkov radiation. These confirmations strengthen the 
motivation to use this effect to search for cascades
induced by predicted diffuse backgrounds of high energy neutrinos,
which are associated with the presence of $\geq 10^{20}$ eV
cosmic rays in many models.
At neutrino energies of about 100 EeV (1 EeV = $10^{18}$ eV),  
cascades in the upper
10 m of the radio-transparent lunar regolith
result in pulses that are detectable
by large radio telescopes at earth~\cite{Zhe88,Dag89}. One
prior experiment has been reported, using the Parkes
64 m telescope~\cite{H96} with about 10 hours of livetime. 

At frequencies above 2 GHz,
ionospheric delay smearing is unimportant, and the signal should appear
as highly linearly-polarized, band-limited 
electromagnetic impulses~\cite{ZHS92,Alv96,Alv97}.
However, since there are many anthropogenic sources of 
impulsive radio emission, the primary problem in
detecting such pulses is eliminating sensitivity to such
interference.

Since 1999 we have been conducting a series of experiments 
to establish techniques to measure such pulses, using the
JPL/NASA Deep Space Network antennas at Goldstone Tracking Facility
near Barstow, California~\cite{Gor99}. We employ the 70 m and 34 m
telecommunication antennas (designated DSS14 and DSS13 respectively)
in a coincidence-type system to solve the problem of 
terrestrial interference, and this approach has proven very
effective. Since mid-2000, the project has moved into a
new status as an ongoing experiment, and receives more 
regularly scheduled observations, subject to the constraints
imposed by the spacecraft telecommunications priorities of the
Goldstone facility. 

Although the total livetime 
accumulated in such an experiment is a relatively small fraction
of what is possible with a dedicated system, the volume of 
material to which we are sensitive, a significant fraction of the
Moon's surface to $\sim 10$ m depth, is enormous, exceeding
100,000 km$^3$ at the highest energies. The resulting 
sensitivity is enough to begin to constrain some
models for diffuse neutrino backgrounds at energies near and
beyond $10^{20}$ eV.
We report on the status of the experiment, and astrophysical
constraints imposed by limits from about 30 hours of livetime.
We are also improving our understanding of the emission
geometry and detection sensitivity through simulations, and
describe initial results in extending our sensitivity to
pulses of lower amplitude.

\section{Description of Experiment}
	
The lunar regolith is an aggregate layer of fine particles and
small rocks, thought to be the accumulated ejecta of meteor impacts
with the lunar surface. It consists mostly of silicates and
related minerals, with meteoritic
iron and titanium compounds at an average level of several per cent, and
traces of meteoritic carbon. It 
has a typical depth range of 10 to 20 m in the maria and valleys, 
but may be hundreds of meters deep in portions of the highlands~\cite{Mor87}.
It has a mean dielectric constant
of $\epsilon \simeq 3$ and a density of $\rho \simeq 1.7$ gm cm$^{-3}$,
both increasing slowly with depth. Measured values for the loss tangent 
vary widely depending on iron and titanium content, but a
mean value at high frequencies is $\tan \delta \simeq 0.003$,
implying a field attenuation length 
at 2 GHz of $(\alpha)^{-1} = 9$~m~\cite{Olh75}.

\subsection{Emission geometry \& Signal Characteristics}

\begin{figure}[b!] 
\centerline{\epsfig{file=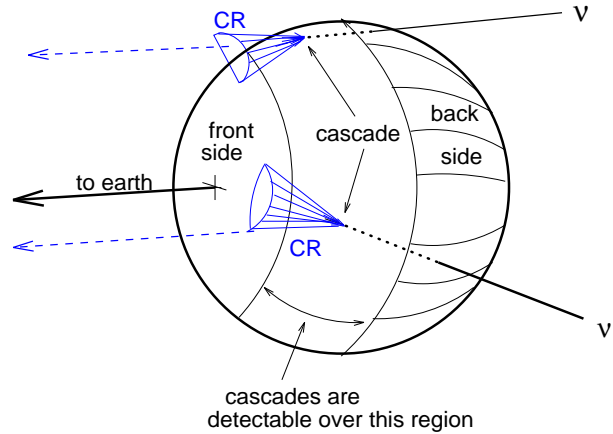,height=2.5in,width=3.5in}}
\vspace{10pt}
\caption{Schematic of the geometry for lunar neutrino cascade event detection.}
\label{moongeom}
\end{figure}

In Fig.~\ref{moongeom} we illustrate the signal emission 
geometry. At 100 EeV the interaction length $L_{int}$ of an electron or
muon neutrino for the dominant deep inelastic hadronic scattering interactions
(averaging over the charged and neutral current processes) is
about 60~km~\cite{Gan00} ($R_{m}=1740$ km). 
Upon interaction, a $\sim 10$ m long
cascade then forms as the secondary particles multiply, and
compton scattering, positron annihilation, and other scattering
processes then lead to a $\sim 20\%$ negative charge excess which
radiates a cone of coherent Cherenkov emission 
at an angle of $56^{\circ}$, with a FWHM 
of $1^{\circ}$. The radiation propagates in the form
of a sub-ns pulse through the regolith to the surface
where it is refracted upon transmission.

Because the angle for total internal reflection (TIR) of
the radiation emitted from the cascade is to first order the
complement of the Cherenkov angle, we consider for the moment
only neutrinos that cascade upon emerging from a penetrating
chord through the lunar limb. Under these conditions the
typical neutrino cascade has an upcoming angle with respect
to the local surface of 
\begin{equation}
\theta_{up} ~=~ \sin^{-1} \left ( {L_{int} \over 2 R_m} \right )
\end{equation}
which implies a mean of $\theta_{up} \sim 1^{\circ}$ at $10^{20}$ eV.

At the regolith surface
the resulting microwave Cherenkov radiation is refracted 
strongly into the forward direction. Scattering from surface
irregularities and demagnification from the interface refraction
gradient fills in the Cherenkov cone, and results in a
larger effective area of the lunar surface over which
events can be detected, as well as a greater acceptance 
solid angle. These effects are discussed in more detail
in section III.A.

\subsection{Antennas \& receivers}

The antennas employed in our search are the shaped-Cassegrainian
70 m antenna DSS14, and the beam waveguide 34 m antenna DSS13, both
part of the NASA Goldstone Deep Space Network (DSN) Tracking Station.
DSS13 is located about 22~km to the SSE of DSS14. The S-band (2.2~GHZ)
right-circular-polarization (RCP) signal from DSS13 is 
filtered to 150~MHz BW, then downconverted with an
intermediate frequency (IF) near 300~MHz.
The band is then further subdivided into high and low frequency
halves of 75 MHz each, and no overlap. These IF signals are then
sent via an analog fiber-optic link to DSS14.
At DSS14, the dual polarization S-band signals are downconverted
with the same 300~MHz IF, and bandwidths of $100$ MHz (RCP)
and 40 MHZ (LCP) are used. A third signal is also employed at
DSS14: a 1.8 GHz (L-band) feed which is off-pointed by $\sim 0.5^{\circ}$
is used as a monitor of terrestrial interference signals; the
signal is downconverted in the same manner as the other signals and
has a 40 MHz bandwidth.

\subsection{Trigger system}

The experimental approach in our initial 12 hours of 
observations was to use a single antenna trigger with
dual antenna data recording~\cite{Gor99}. This was accomplished
by using the local S-band signals as DSS14 to form a 2-fold coincidence
with an active veto from the L-band interference monitor.
Since any system with an active veto is subject to potential
unforeseen impact on the trigger efficiency, we have now
developed an approach which utilizes signals from both antennas
to form a real-time dual-antenna trigger, with no active veto.

\begin{figure}[b!] 
\centerline{\epsfig{file=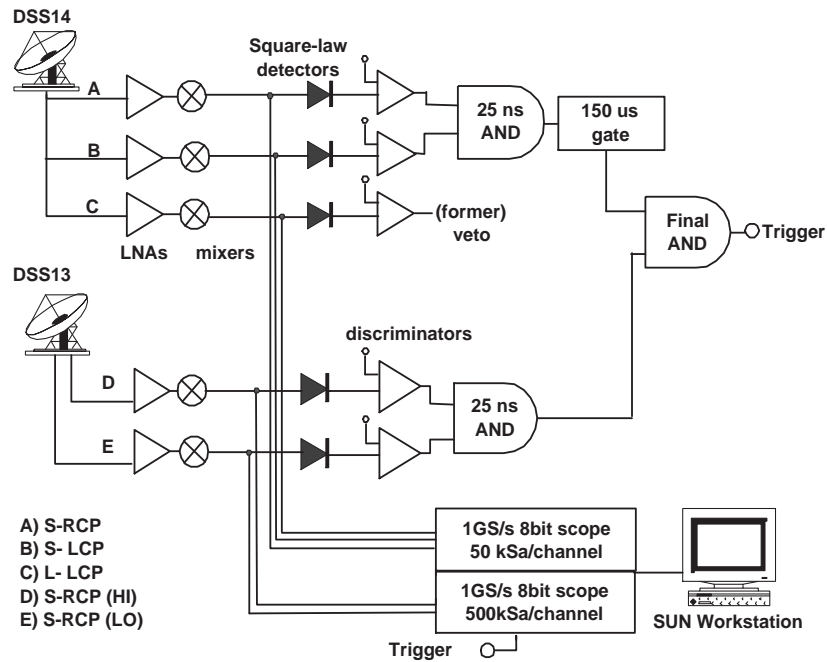,height=3.5in,width=4.5in}}
\vspace{10pt}
\caption{The GLUE trigger system used for the lunar neutrino search.}
\label{trigger1}
\end{figure}

Fig.~\ref{trigger1} shows the layout of the trigger. The four
triggering signals from the two antennas are converted to
unipolar pulses using tunnel-diode square-law detectors.
Stanford Research Systems SR400 discriminators are used for the
initial threshold level, and these are set to maintain
a roughly constant singles rate, typically 0.5-1 kHz/chan for
DSS14 and 30 kHz/chan for DSS13 (DSS13's rate is higher due to a
lower threshold, compensating for the reduced aperture size). 
A local coincidence is then
formed for each antenna's signals. The DSS14 coincidence 
between both circular polarizations ensures that the signals
are highly linearly polarized, and the DSS13 coincidence
helps to ensure that the signal is broadband.

\begin{figure}[b!] 
\centerline{\epsfig{file=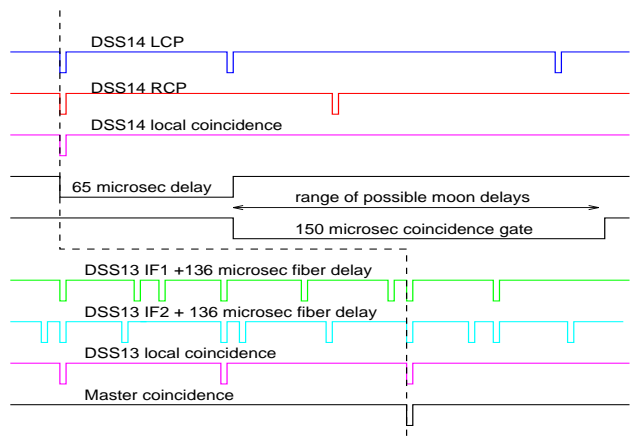,height=2.5in,width=3.5in}}
\vspace{10pt}
\caption{A timing diagram for the GLUE trigger system.}
\label{trigger2}
\end{figure}

Fig.~\ref{trigger2} indicates the timing sequence for a trigger
to form (negative logic levels are used here). 
A local coincidence 
at DSS14, typically with a 25~ns gate, initiates the trigger
sequence. After a 65~$\mu$s delay, a 150~$\mu$s gate is opened
(the delays compensate for the 136~$\mu$s fiber delay between
the two antennas).
This large time window encompasses the possible geometric delay range for
the moon throughout the year. Use of a smaller window is possible
but would require delay tracking and a thus more stringent need for
testing and reliability; use of a large window avoids 
this and a tighter coincidence can then be required offline.

If a 25~ns local coincidence now forms between the two DSS13 signals within the
allowed 150~$\mu$s window, a trigger is formed. The sampling scopes are
then triggered, and a 250~$\mu$s record, sampled at
1 Gs/s, is stored. The average trigger rate, due primarily
to random coincidences of thermal noise fluctuations, is about
1.6 mHz, or 1 trigger every 5 minutes or so. Terrestrial
interference triggers are uncommon (a few percent of the total), 
but can occasionally increase in number when a large burst
of interference occurs at either antenna, with DSS14 more sensitive
to this effect. The deadtime per event is about 6~s; thus
on average we maintain about 99\% livetime during a run.

\section{Estimated sensitivity}

Estimates of the sensitivity of radio telescope observations
usually involve systems that integrate total power for some
time constant $\Delta t$ which is in general much longer
than the antenna's single temporal mode duration which
is given by the inverse of the bandwidth: $\tau = (\Delta \nu)^{-1}$. 
Since the pulses of interest in our experiment are much shorter
than this time scale, the observed pulse
structure of induced voltage in the antenna receiver is determined
only by the bandpass function; that is, the pulses are
band-limited. Thus the typical dependence of sensitivity
on the factor $\sqrt{\Delta t \Delta \nu}$ does not
obtain; this factor is always unity in band-limited pulse
detection.

Because much of the theoretical work in describing such pulses
has been done in terms of field strength rather than
power, we analyze our sensitivity in these terms as well.
Such analysis is also compatible with the receiving 
system, which records antenna voltages proportional to the
incident electric field, and leads to a more linear analysis.
It also yields signal-to-noise ratio estimates which are
consistent with Gaussian statistics, since thermal noise
voltages are described by a Gaussian random process.

The expected field strength per unit bandwidth 
from a cascade of total energy $W_T$
can be expressed as~\cite{ZHS92,Alv96,Alv97}: 
\begin{equation}
\label{zhs}
E_0~ ({\rm V~m^{-1}~MHz^{-1}} ) ~=~ {2.53 \times 10^{-7} \over R }
\left ( {W_{T} \over 1~{\rm TeV}} \right )
{\nu \over \nu_0} \left ( {1 \over 1 + 0.4(\nu/\nu_0)^{1.44} } \right ) ~,
\end{equation}
where  $R$ is the distance to the source in m, 
$\nu$ is the radio frequency, and 
the decoherence frequency is $\nu_0 \simeq 2500$ MHz for regolith material
($\nu_0$ scales mainly by radiation length). For typical
parameters in our experiment, a $10^{19}$ eV cascade will
result in a peak field strength at earth of $E \simeq 0.5~\mu$V m$^{-1}$
for a 70 MHz BW.
Equation~\ref{zhs} has now been verified to within factors of $2$
through accelerator tests~\cite{Gor00,Sal01} using silica sand targets
and $\gamma$-ray-bunch-induced cascades with $W_T \leq 10^{19}$ eV per bunch.

Given that the use of a dual antenna trigger has virtually eliminated
the problem of terrestrial interference that was the primary
limitation to the sensitivity of the one previous experiment~\cite{H96},
we can now express the minimum detectable field strength $E_{min}$ for each
antenna in terms of the induced signal and the 
thermal noise background. 

The expected
signal strength $E_0$ induces a voltage at the antenna receiver given 
by
\begin{equation}
v_s ~=~ h_e E_0 \Delta \nu
\end{equation}
where the antenna effective height $h_e$ is given by~\cite{Kra88}
\begin{equation}
h_e ~=~ 2 \sqrt{ { Z_a \eta A \over Z_0 } } \cos \theta_p
\end{equation}
where $Z_a$ is the antenna radiation resistance, 
$\eta$ and $A$ are the antenna efficiency and area, respectively,
$Z_0=377~\Omega$ is the impedance of free space, and $\theta_p$ the 
polarization angle of the antenna with respect to the plane of
polarization of the radiation.

The average thermal noise voltage in the system is given by
\begin{equation}
v_n ~=~ \sqrt{4 k T_{sys} Z_T \Delta \nu}~.
\end{equation}
Here $k$ is Boltzmann's constant, $T_{sys}$ is the system 
thermal noise temperature,
and $Z_T$ the termination impedance of the receiver.
If we assume that $Z_a \approx Z_T$ then the resulting
SNR $S$ is
\begin{equation}
S ~\equiv~ {v_s \over v_n} ~=~ 
E_0 \cos \theta_p \sqrt{{\eta A \Delta \nu \over k T_{sys} Z_0 }}~.
\end{equation}
The minimum detectable field strength ($E_0 \rightarrow E_{min}$)
is then given by
\begin{equation}
E_{min} ~=~ S \sqrt{{k T_{sys} Z_0 \over \eta A \Delta \nu}} 
~{1 \over \cos \theta_p}~.
\end{equation}
Combining this with equation~\ref{zhs} above, the threshold
energy for pulse detection is
\begin{equation}
W_{thr} (\mbox{EeV}) ~\simeq~ 4.0  
\left ( {E_{min} \over {\rm 1 V~m^{-1} MHz^{-1}} } \right )
\left ( {R \over {\rm 1~m}} \right )
 {\nu_0 \over \nu} 
\left [ {1 + 0.4 \left ( {\nu \over \nu_0} \right )^{1.44} } \right ] ~.
\end{equation}

For the lunar observations on the limb, which make up about 85\% of
the data reported here, $T_{sys} \simeq 110$ K, $\nu = 2.2$ GHz,
and the average $\Delta \nu \simeq 70$ MHz. For the 70 m antenna,
with efficiency $\eta \simeq 0.8$, the minimum detectable
field strength is $E_{min} \simeq 1.2 \times 10^{-8}$ V m$^{-1}$ MHz$^{-1}$
for $\cos \theta_p = 0.7$. The estimated threshold energy for
these parameters is $W_{thr} = 2.8 \times 10^{19}$ eV, assuming a
detection level of $S=5 ~(5 \sigma)$ per IF at DSS14 (with a somewhat
lower requirement at DSS13 in coincidence).

\subsection{Monte Carlo results}

To estimate the effective volume and acceptance solid angle
as a function of incoming neutrino energy, 
events were generated at discrete neutrino energies, including
the current best estimates of both charged and neutral current
cross sections~\cite{Gan00}, and the Bjorken-y distribution.
Both electron and muon neutrino interactions were included,
and Landau-Pomeranchuk-Migdal effects in the shower formation
were estimated~\cite{Alv97}. At each
neutrino energy, a distribution of cascade angles and depths with respect
to the local surface was obtained, and a refraction propagation
of the predicted Cherenkov angular distribution
was made through the regolith surface, including absorption and reflection
losses and a first
order roughness model. Antenna thermal noise fluctuations were
included in the detection process.

\begin{figure}[h!] 
\centerline{\epsfig{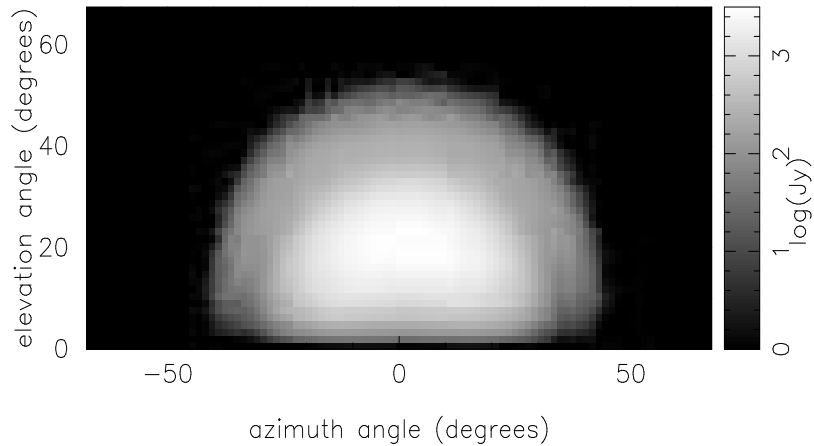}}
\vspace{10pt}
\caption{The microwave Cherenkov radiation pattern from an event
in the lunar regolith.}
\label{CRpattern}
\end{figure}

A portion of the simulation is shown in Fig.~\ref{CRpattern}.
Here the flux density is shown as it would appear projected on the sky,
with (0,0) corresponding to the tangent to the lunar surface
in the direction of the original cascade. The units are Jy 
(1 Jy = $10^{-26}$ W m$^{-2}$ Hz$^{-1}$) as measured at earth,
and the plot is an average over several hundred events at different
depths and a range of $\theta_{up}$ consistent with a $10^{20}$ eV
neutrino interaction, averaging over inelasticity effects and a mixture
of electron and muon neutrinos consistent with 
decays from a hadronic $\pi^{\pm}$ source. 

Although the averaging has 
broadened the distribution somewhat, a typical cascade
still produces a flux density pattern of comparable angular size.
The angular width of the pattern directly
increases the acceptance solid angle, and the angular
height increases the annular
band of the lunar surface over which neutrino events can be detected,
as indicated in Fig.~\ref{moongeom}. The net effect is that,
although the specific flux density of the events are lowered somewhat by
refraction and scattering, the effective volume and acceptance
solid angle are significantly increased. The 
neutrino acceptance solid angle, in particular,
is about a factor of 50 larger than the apparent solid angle of the 
moon itself.

\subsection{EHE cosmic rays}
We have noted above that the refraction geometry of the
regolith favors emission from cascades that are upcoming
relative to the local regolith surface. Thus to first order
EHE cosmic ray events, which cascade within a few tens of cm
as they enter the regolith, will not produce detectable
pulses since their emission with be totally internally
reflected within the regolith. This effect has been
now demonstrated in an accelerator experiment~\cite{Sal01}.

\begin{figure}[h!] 
\centerline{\epsfig{file=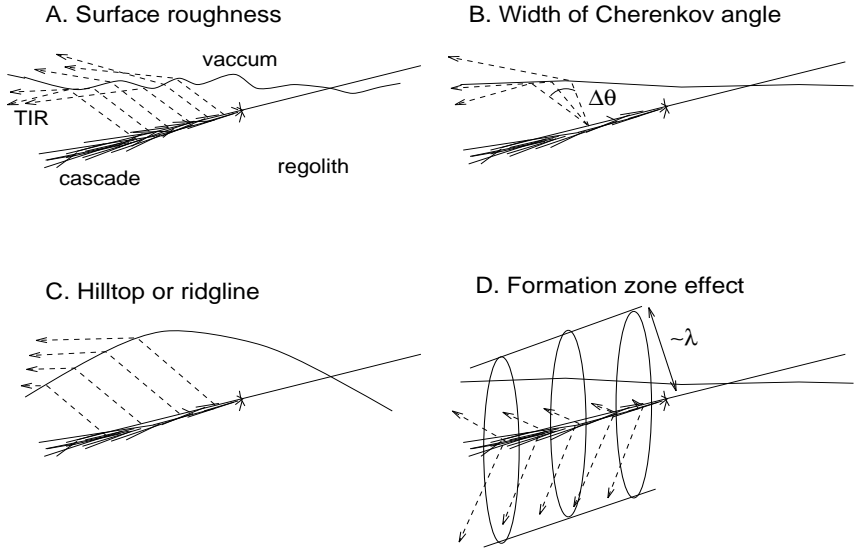,height=3in,width=4.5in}}
\vspace{10pt}
\caption{Various effects associated with EHE cosmic ray hadron interactions.}
\label{ehecr}
\end{figure}

This conclusion does not account for several effects however.
These effects are illustrated in Fig.~\ref{ehecr}.
In Fig.~\ref{ehecr}A, 
the varying surface angles due surface roughness on scales greater than
a wavelength will lead to escape of some radiation
from cosmic ray cascades.
Fig.~\ref{ehecr}B illustrates that, because the 
Cherenkov angular distribution is not infinitely narrow
arround the Cherenkov angle (FWHM $\sim 1^{\circ}$), 
emission at angles larger then
the Cherenkov angle can escape total internal reflectance.
In Fig.~\ref{ehecr}C cascades from cosmic rays that enter along ridgelines can
encounter a change in slope of the local surface that results
in more efficient transmission of the radiation.

Even if total internal reflection strongly suppresses detection
of cosmic rays in cases A and B, the latter case C of favorable
surface geometry along ridgelines or hilltops will lead to
some background of EHE cosmic ray events. We have not as yet
made estimates of this background.\footnote{These conclusions
also apply of course to the fraction of neutrinos which interact
on entering the regolith as well as those which interact 
near their projected exit point. Thus we have not yet accounted
for all of the possible neutrino events as well as the cosmic ray
background events.}

Fig.~\ref{ehecr}D shows the formation zone aspect of the process
of Cherenkov emission from near-surface cosmic ray cascades.
This constraint may suppress Cherenkov production
even if surface roughness and the width of the
Cherenkov distribution would otherwise favor some escape of emission.

It has now been conclusively shown~\cite{Tak00} that coherent
Cherenkov emission is a process involving the bulk dielectric
properties of the radiating material. Cherenkov radiation
is induced over a macroscopic region of the dielectric
(with respect to the scale of a wavelength), and does not even require
that the charged particles enter the dielectric for radiation
to be produced---a proximity of several wavelengths or less
is sufficient~\cite{Ulr66}. A corollary to this result is that a cascade
travelling along very near a boundary of the dielectric will not
radiate (or radiate only weakly) into the hemisphere with the boundary.

Thus in the case of a cosmic ray entering the regolith at near
grazing incidence (say within $\sim 1^{\circ}$) the resulting
cascade reaches maximum  within $\sim 3$ cm of the surface,
still less than a wavelength for S-band observations. We
therefore expect that the suppression of Cherenkov emission
in such events significantly 
reduces our sensitivity to cosmic rays. Such effects have not
been included yet in other estimates~\cite{Alv96,Alv01} of the
cosmic ray detection efficieny of such experiments.

\section{Results}

\begin{figure}[h!] 
\centerline{\epsfig{file=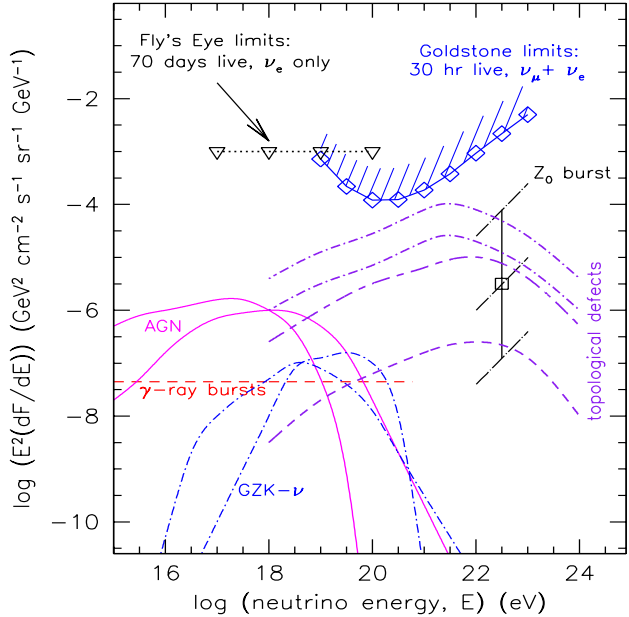,height=3.5in,width=4in}}
\vspace{10pt}
\caption{Plot of model neutrino fluxes and limits from the Fly's
Eye experiment and the present work.}
\label{limits}
\end{figure}

Figure~\ref{limits} plots the predicted fluxes of 
EHE neutrinos from a number of
models including AGN production~\cite{Man96}
gamma-ray bursts~\cite{BW99}, 
EHE cosmic-ray interactions~\cite{HS85},
topological defects~\cite{Yos97,Bha92}, and
the $Z_0$ burst scenario~\cite{Wei99}. 
Also plotted are limits from about 70 days of Fly's Eye 
livetime~\cite{Bal85} (accumulated in several years of
runtime), which apply only to electron neutrino events.

Our initial 90\% CL limit, for 30 hours of livetime
is shown plotted with diamonds (see also Table 1), based
on the observation of no events above an equivalent $5\sigma$ level
amplitude (referenced to the 70 m antenna) consistent with 
the direction of the moon. These limits assume a monoenergetic
signal at each energy; thus they are differential limits
and independent of source spectral
model, and represent the most conservative limits we can apply.
Our limits just begin to constrain the highest
topological defect model~\cite{Yos97} for which 
we expected a total of order 1--2 events. 

\begin{table}[h]
\centering
\caption{Differential limits on mono-energetic EHE neutrino fluxes.}
\label{nulimits}
\medskip
\centering
\begin{tabular}{lccccccccc}
\tableline
Energy (eV)& $10^{19}$ & $3 \times 10^{19}$ &$10^{20}$ &$3 \times 10^{20}$ &$10^{21}$ & $3 \times 10^{21}$ & $10^{22}$ & $3 \times 10^{22}$ & $10^{23}$ \\
\tableline \\
$\log_{10}(E^2 dF/dE)$ & -3.14 & -3.66 & -3.92 & -3.91 & -3.73 & -3.42 & 
-3.03 & -2.66 & -2.30 \\
 (GeV cm$^{-2}$ s$^{-1}$ sr$^{-1}$)  & & & & & & & &\\
\tableline
\end{tabular}
\end{table}

\begin{figure}[ht!] 
\centerline{\epsfig{file=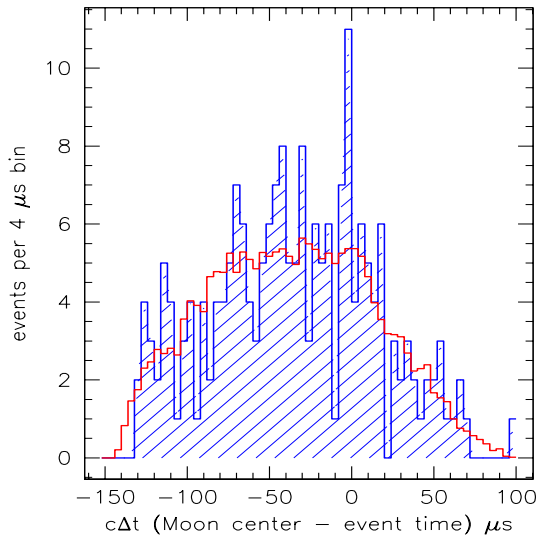,height=3in,width=3.5in}}
\vspace{10pt}
\caption{Low amplitude event histogram of GLUE data in delay with respect
to the moon center delay. The overlain histogram is the
expected background level. An excess occurs near zero delay, with a
2 microsecond offset.}
\label{smallev}
\end{figure}

In addition to the limits set above from the non-observation
of events above, we have also analyzed events which triggered the
system, but did not pass our more stringent software amplitude
cuts. A sample of events was prepared by applying our standard
cuts to remove terrestrial interference events. We required
somewhat tighter timing that the hardware trigger, as well
as band-limited pulse shape, but allowed smaller amplitudes,
typically corresponding to $\sim 4.5 \sigma$ at DSS14, and
about $3 \sigma$ at DSS13.
The results are shown in Fig.~\ref{smallev}, where the
passing events have been binned according to their delay timing
with respect to the expected delay from an event at the center
of the moon. The background level (solid line) has been determined by
randomizing the UT of the events and indicates the somewhat
non-uniform seasonal coverage of our observations.

An excess is observed in the vicinity of zero delay where the
lunar events are expected to cluster. At present there 
is a $\sim 2 \mu$s offset from zero delay; this is too 
large to be accounted for by differential delays to the
lunar limb, which can produce offsets of several hundred ns. 
Further study of the low amplitude events is in progress.

\section{Conclusions}

We have developed a robust system for observing microwave
pulses produced in the lunar regolith by electromagnetic
particle cascades above $\sim 10^{19}$ eV. We have
operated this system to achieve a livetime of 30 hours, with
no large apparent signals detected to date. We have set
conservative upper limits on the diffuse cosmic neutrino fluxes
over the energy range from $10^{19-23}$ eV.
We have also begun to analyze smaller events and have
some preliminary indications that a signal may be present,
but requiring further study.

\vspace{0.5cm}

\small{We thank Michael Klein, George Resch, and the 
staff at Goldstone for their enthusiastic support of our efforts.
This work was performed in part
at the Jet Propulsion Laboratory, California Institute of Technology, under 
contract with NASA, and supported in part by the Caltech President's
Fund,  by DOE contract DE-FG03-91ER40662 at UCLA, the Sloan Foundation,
and the National Science Foundation.
}

\end{document}